\documentclass[slac_one,nofootinbib]{revtex4-2}
\usepackage{graphicx}
\usepackage{fancyhdr}
\begin{document}

\title{Trust but verify: The case for astrophysical black holes}

\author{Scott A. Hughes}
\affiliation{Department of Physics and MIT Kavli Institute, 77
Massachusetts Avenue, Cambridge, MA 02139}

\begin{abstract}
This article is based on a pair of lectures given at the 2005 SLAC
Summer Institute.  Our goal is to motivate why most physicists and
astrophysicists accept the hypothesis that the most massive, compact
objects seen in many astrophysical systems are described by the black
hole solutions of general relativity.  We describe the nature of the
most important black hole solutions, the Schwarzschild and the Kerr
solutions.  We discuss gravitational collapse and stability in order
to motivate why such objects are the most likely outcome of realistic
astrophysical collapse processes.  Finally, we discuss some of the
observations which --- so far at least --- are totally consistent with
this viewpoint, and describe planned tests and observations which have
the potential to falsify the black hole hypothesis, or sharpen still
further the consistency of data with theory.
\end{abstract}

\maketitle

\thispagestyle{fancy}

\section{Background}

Black holes are among the most fascinating and counterintuitive
objects predicted by modern physical theory.  Their counterintuitive
nature comes not from obtuse features of gravitation physics, but
rather from the extreme limit of well-understood and well-observed
features --- the bending of light and the redshifting of clocks by
gravity.  This limit is inarguably strange, and drives us to
predictions that may reasonably be considered bothersome.  Given the
lack of direct observational or experimental data about gravity in the
relevant regime, a certain skepticism about the black hole hypothesis
is perhaps reasonable.  Indeed, one can understand Eddington's
somewhat plaintive cry ``I think there should be a law of Nature to
prevent a star from behaving in this absurd way!''  {\cite{bhtw}}.

To the best of our knowledge, there is {\it no} such law of Nature.
We expect that in any object which is sufficiently massive, gravity
can overwhelm all opposing forces, causing that object to collapse
into a black hole --- a region of spacetime in which gravity is so
intense that not even light can escape.  Indeed, astronomical
observations have now provided us with a large number of {\it
candidate} black holes --- dark objects which are so massive and
compact that the hypothesis that they are the black holes of general
relativity (GR) is quite plausible.  These objects are now believed to
be quite common --- extremely massive black hole candidates (several
$10^5$ solar masses to $\sim 10^9$ solar masses) reside in the cores
of almost all galaxies, including our own; stellar mass candidates
(several to several tens of solar masses) are seen in orbit with stars
in our galaxy; and evidence suggests that ``intermediate'' mass black
holes may fill the gap between these limits.  Not only is there no law
preventing stars from ``behaving in this absurd way'', Nature appears
to revel in such absurdities.

It must be emphasized that, from the purely observational perspective,
at present these objects are black hole {\it candidates}.  We have not
unequivocally established that {\it any} of these objects exhibits all
of the properties that GR predicts for ``real'' black holes.  GR is
consistent with Newton's theory of gravity in the appropriate limit.
This means that, for example, orbits of black holes are totally
consistent with Newtonian gravity except for strong field (small
radius) orbits.  Establishing whether these objects are the black
holes of GR requires observations that probe close to the black hole
itself.  Because black holes are so compact, this is a challenging
proposition.  Consider the putative black hole at the core of our
galaxy.  If it is a black hole as described by GR, its radius should
be about $10^7$ kilometers.  It is located roughly 7500 parsecs from
the earth.  The black hole thus subtends an angle $\delta\theta \sim
10^7\,{\rm km}/(7.5\times 10^3\,{\rm pc}) \sim 0.01$ milliarcseconds
on the sky!  Resolving such small scale phenomena is extraordinarily
difficult.

Given that no observations have unequivocally established that black
hole candidates are GR's black holes, it is not surprising that
alternate hypotheses have been advanced to describe these candidates.
Since GR is a classical theory and Nature is fundamentally quantum,
one might {\it expect} something other than classical GR to describe
black hole candidates.  Such a modification certainly seems necessary
in order to, for example, resolve black hole information paradoxes
{\cite{info}}.

The standard view is that, though modifications to classical GR are
needed, their impact is most likely negligible for astrophysical black
holes.  Black holes describe gravitational effects far beyond what we
normally encounter, and so appear to be ``extreme objects''.  Within
the context of general relativity, however, they are not so extreme.
In GR, the ``fundamental'' gravitational field is not manifested as
acceleration; instead, it is the {\it spacetime curvature}, which is
manifested as tidal fields.  Though black holes have enormous
gravitational acceleration, their tides can be quite gentle.  In
particular, astrophysical black hole tides are {\it far} smaller than
the level at which one would expect quantum corrections to be
important.

Regardless of whether one accepts this standard view of modifications
to GR theory or whether one postulates more drastic modifications, one
needs some kind of ``gold standard'' against which to compare
observations.  The view we take here is that, given the number of
possible ways in which one can modify classical GR, it is currently
most reasonable to treat the classical black hole solution as our gold
standard.  {\it If} observations show that this is inadequate, then we
would have an experimentally motivated direction to pursue
modifications to general relativity and to black holes.

The goal of these lectures is to summarize the case for black holes
within classical GR.  We first discuss, in Sec.\ {\ref{sec:schw}}, the
simplest, exact black hole solutions of general relativity.  We
illustrate that they have a completely reasonable, non-singular nature
(at least in their exteriors!), but that uncovering this nature can be
a bit subtle owing to some pathologies in the coordinates that are
commonly used to describe them.  Changing to coordinates which
eliminate these pathologies, we show that light can never escape once
inside a certain radius (the ``black''), and thus any object which
falls inside is doomed to fall until diverging tidal forces tear it
apart (the ``hole'').

The solution we use to build this description corresponds to a black
hole that exists for all time.  One might worry that the features we
discovered using it are a contrivance of that artificial, eternal
solution.  Can we really create a region of spacetime with these
features beginning with well-behaved ``normal'' matter?  As shown in
Sec.\ {\ref{sec:collapse}}, we can.  A well-known solution, developed
by Oppenheimer and his student Snyder in 1939 {\cite{os39}} shows {\it
analytically} that pressureless, spherical matter quickly collapses
from a smooth, well-behaved initial distribution to form a black hole
with {\it exactly} the properties of the exact solution that we
initially discussed.  Numerical calculations show that including
pressure and eliminating spherical symmetry leads to the same result.

These results depend on the solution having no angular momentum.  Roy
Kerr lifted this requirement when he discovered the solution which now
bears his name {\cite{kerr}}.  Spinning black holes are the generic
case that we expect astrophysically; as discussed in Secs.\
{\ref{sec:kerr}} and {\ref{sec:unique_stable}} classical general
relativity predicts the Kerr black hole to be the unique and stable
endpoint of gravitational collapse.  We discuss the perturbative
techniques which were first used to advance this result, and discuss
numerical calculations which show that it holds generically.

The Kerr black hole is thus the object which general relativity
predicts describes the many black hole candidates which we observe
today.  What remains is to test this prediction.  In Sec.\
{\ref{sec:test}}, we briefly describe some of the observations being
developed today which can probe into the strong field of these objects
and confront the Kerr black hole hypothesis with data.

Throughout this article, we assume that the reader has basic
familiarity with general relativity; in particular, the notion of
geodesics and the Einstein field equations will play an important
role.  Carroll's SLAC Summer Institute lectures covered this material;
the reader is referred to those lectures for GR background.  The
presentation of Carroll's textbook {\cite{carroll}} and the classic
textbook by Misner, Thorne, and Wheeler (MTW) {\cite{mtw}} influenced
the development of these lectures quite a bit.  Readers interested in
a very detailed discussion of black hole physics would be well-served
with Poisson's textbook {\cite{poisson}}.

In most of the article, we use ``relativist's units'' in which $G = 1
= c$.  These units nicely highlight the fact that general relativity
has no intrinsic scale; as we shall see, all important lengthscales
and timescales are set by and become proportional to black hole mass.
We shall sometimes restore $G$s and $c$s to make certain quantities
more readable.  Useful conversion factors for discussing astrophysical
black holes come from expressing the solar mass ($M_\odot$) in these
units:
\begin{eqnarray}
M_\odot &\to& \frac{GM_\odot}{c^2} = 1.47663\,{\rm km} \simeq 1.5\,{\rm
 km}
\\
&\to& \frac{GM_\odot}{c^3} = 4.92549\times 10^{-6}\,{\rm sec} \simeq
5\,\mu{\rm sec}\;.
\end{eqnarray}

\section{The Schwarzschild black hole}
\label{sec:schw}

The simplest black hole solution is described by the {\it
Schwarzschild metric}, discovered by Karl Schwarzschild in 1916
{\cite{schw16}}.  In what are now known as ``Schwarzschild
coordinates'', this metric takes the form
\begin{equation}
ds^2 = -\left(1 - \frac{2M}{r}\right)dt^2 + \frac{dr^2}{1 - 2M/r} +
r^2\left(d\theta^2 + \sin^2\theta\,d\phi^2\right)\;.
\label{eq:schw_metric}
\end{equation}
By building the Einstein tensor from this metric, it is a simple
matter to show that the Schwarzschild solution is an exact solution to
the Einstein field equations $G_{ab} = 8\pi T_{ab}$ for $T_{ab} = 0$
--- i.e., it's a vacuum solution.

Since it satisfies Einstein's equations, we can be satisfied that this
metric is mathematically consistent.  Is it however a physically
meaningful solution?  Is this metric one that we might use to describe
real objects in the universe, or is it too idealized to be worthwhile?
To answer this, we first examine several limiting regimes of this
spacetime.  First consider $r \gg M$:
\begin{equation}
ds^2 \simeq -\left(1 - \frac{2M}{r}\right)dt^2 + \left(1 +
\frac{2M}{r}\right)dr^2 + r^2\left(d\theta^2 +
\sin^2\theta\,d\phi^2\right)\;.
\end{equation}
Examine slow motion geodesics (spatial velocity $|dx^i/dt| \ll c$) in
this spacetime.  The equation of motion which we find describes the
coordinate motion of bodies reduces, in this limit, to
\begin{equation}
\frac{d^2x^i}{dt^2} = -\frac{Mx^i}{r^3}\;.
\end{equation}
In other words, the $r \gg M$ limit of the Schwarzschild metric
reproduces Newtonian gravity!  Clearly, the Schwarzschild metric
represents the external ``gravitational field'' of a monopolar mass
$M$ in general relativity\footnote{This limiting behavior is used to
derive the Schwarzschild metric from first principles.  Beginning with
the most general static, spherically symmetric line element, $ds^2 =
-e^{2\Phi(r)}dt^2 + e^{2\Lambda(r)}dr^2 + R(r)^2(d\theta^2 +
\sin^2\theta\,d\phi^2)$, one enforces the vacuum Einstein equations
and finds $R(r) = r$, $\Lambda(r) = -\Phi(r)$.  Correspondence with
the Newtonian limit leads to $e^{2\Phi(r)} = 1 - 2M/r$.}.

Let us now study this metric in this coordinate systems more
carefully.  First, consider a ``slice'' of the spacetime with constant
$r$ and constant $t$:
\begin{equation}
ds^2 = r^2\left(d\theta^2 + \sin^2\theta\,d\phi^2\right)\;.
\label{eq:sphere}
\end{equation}
This is the metric that is used to measure the distance between points
on the surface of a sphere; in other words, this is a spherically
symmetric spacetime, with $\theta$ and $\phi$ chosen to be the
``usual'' spherical angles that cover the sphere.  More interestingly,
this construction helps us to understand the meaning of the coordinate
$r$: Eq.\ (\ref{eq:sphere}) shows us that the surface area of this
sphere is just $4\pi r^2$ --- just as our intuition would lead us to
expect.

Now, allow $r$ to vary:
\begin{equation}
ds^2 = \frac{dr^2}{1 - 2M/r} + r^2\left(d\theta^2 +
\sin^2\theta\,d\phi^2\right)\;.
\label{eq:vary_r}
\end{equation}
The distance between two spatial points $(r_1,\theta,\phi)$ and
$(r_2,\theta,\phi)$ is {\it not} just $\Delta r = r_2 - r_1$ as our
intuition would have led us to believe!  Instead we must integrate
the line element,
\begin{equation}
{\rm Distance} = \int_{r_1}^{r_2} \frac{dr}{\sqrt{1 - 2M/r}} =
 \frac{r_2 - 2M}{\sqrt{1 - 2M/r_2}} - \frac{r_1 - 2M}{\sqrt{1 -
 2M/r_1}} + 2M\ln\left[\sqrt{\frac{r_2}{r_1}}\left(\frac{1 + \sqrt{1 -
 2M/r_2}}{1 + \sqrt{1 - 2M/r_1}}\right)\right]\;.
\end{equation}
The coordinate $r$ is an ``areal'' radius --- it labels spherical
surfaces of area $4\pi r^2$, but it does {\it not} label proper
distance in a simple way.

Finally, we need to understand the meaning of the timelike coordinate
$t$.  When we are {\it extremely} far from the mass $M$, the line
element (\ref{eq:schw_metric}) reduces to that of flat spacetime:
\begin{equation}
ds^2 \approx -dt^2 + dr^2 + r^2\left(d\theta^2 +
\sin^2\theta\,d\phi^2\right)\;.
\end{equation}
The $t$ coordinate is clearly time as measured by observers in this
``asymptotically flat'' region.  This is the meaning of $t$: It is the
label that {\it very distant observers} use to measure the passing of
time.  This is a very convenient time label for many purposes, but is
not so good for many others.  For example, it is a superb coordinate
for describing features that would be measured by distant observers
watching processes near a black hole.  It is an absolutely lousy
coordinate for describing what occurs to an object that falls into a
black hole.  Many misunderstandings about the nature of black holes
are fundamentally due to the nature of the Schwarzschild time
coordinate.

Having examined and explained the nature of the coordinates with which
we have presented the Schwarzschild solution, let us now begin to
understand the spacetime itself.  It is clear that there are some
oddities we should worry about --- some metric components diverge or
go to zero at $r = 2M$; all metric components show funny behavior as
$r \to 0$.  As we shall see, the first behavior reflects a coordinate
singularity --- the spacetime is perfectly well behaved at $r = 2M$,
but the coordinates are not.  The second pathology by contrast
reflects a real singularity --- a divergence in tidal forces,
reflecting a breakdown in the classical solution.

\subsection{Birkhoff's Theorem}

In deciding how we should regard the Schwarzschild solution, perhaps
the most important result to bear in mind is Birkhoff's Theorem:
\begin{quote}
The exterior spacetime of {\it all} spherical gravitating bodies is
described by the Schwarzschild metric.
\end{quote}
The Schwarzschild metric describes a portion of the spacetime of {\it
any} spherically symmetric body.  Consider for example a spherical
star of radius $R_*$.  Birkhoff's theorem then guarantees that Eq.\
(\ref{eq:schw_metric}) describes the star's spacetime for $r > R_*$.
(Obviously, some other metric must describe the spacetime in the
star's interior, $r < R_*$, since the interior contains matter.)

In fact, Birkhoff's theorem is even more powerful than this: It
applies even if the spacetime is {\it time dependent}, so long as the
time dependence is spherical in nature.  Thus, the exterior spacetime
of a star that undergoes radial pulsations likewise is described by
the Schwarzschild metric.  This tells that spherical oscillations
cannot affect the star's exterior gravitational field\footnote{In
retrospect, this makes a lot of sense: The only way to causally affect
the exterior field is for information about changes to the star's
gravity to propagate radiatively.  Monopolar oscillations cannot
radiate in general relativity, just as they don't radiate in
electrodynamics.}.

Birkhoff's Theorem is in fact quite simple to prove.  Carroll
{\cite{carroll}} gives a very careful proof; here, we outline a
simpler proof, based on that given in MTW {\cite{mtw}}.  We begin by
writing down the general line element for a time dependent, spherical
spacetime:
\begin{equation}
ds^2 = -a(r,t)^2\,dt^2 - 2 a(r,t)b(r,t)\,dr\,dt + d(r,t)^2\,dr^2 +
r^2\left(d\theta^2 + \sin^2\theta\,d\phi^2\right)\;.
\label{eq:birkhoff1}
\end{equation}
[Without loss of generality, we have chosen the coefficient of the
angular sector to be $r^2$ rather than some function $R(r)^2$.  By
doing so, we are choosing our coordinate $r$ to be an areal radius.]
We can immediately simplify this by changing our time coordinate:
define
\begin{equation}
e^{\Phi(r,t')} dt' = a(r,t)\,dt + b(r,t)\,dr\;.
\end{equation}
Insert this into (\ref{eq:birkhoff1}), define $e^{2\Lambda(r,t')} =
b^2(r,t) + d^2(r,t)$, and drop the primes on $t$ in the result.  We
find
\begin{equation}
ds^2 = -e^{2\Phi(r,t)}\,dt^2 + e^{2\Lambda(r,t)}\,dr^2 +
r^2\left(d\theta^2 + \sin^2\theta\,d\phi^2\right)\;.
\label{eq:birkhoff2}
\end{equation}
We now enforce the vacuum Einstein equations: Compute the Einstein
tensor $G_{ab}$ and equate it to $0$.  (We see here why Birkhoff
applies to a gravitating body's exterior; on the interior, we would
set the right hand side to something appropriate to the body's
interior matter distribution.)  We find several differential equations
that constrain the functions $\Phi(r,t)$ and $\Lambda(r,t)$.  For
example, the $r,t$ component of the Einstein tensor yields [cf.\ MTW,
Eq.\ (32.3b)]
\begin{equation}
\frac{2e^{-(\Phi + \Lambda)}}{r}\frac{\partial\Lambda}{\partial t} = 0
\qquad\longrightarrow\qquad \frac{\partial\Lambda}{\partial t} = 0\;.
\end{equation}
$\Lambda$ is a function of $r$ only --- it has no time dependence.
When this is taken into account, the components of the Einstein tensor
that determine $\Lambda$ are identical to the static case, and we find
\begin{equation}
e^{2\Lambda} = \frac{1}{1 - 2M/r}\;,
\end{equation}
just as before.  The remaining Einstein tensor components constrain
the function $\Phi(r,t)$; enforcing the vacuum Einstein equation gives
us
\begin{equation}
e^{2\Phi} = \left(1 - 2M/r\right)e^{2f(t)}\;,
\end{equation}
where $f(t)$ is an arbitrary function of time (arising as an
integration constant).  It can be eliminated by changing time
variables once again: $dt' = e^{f(t)}\,dt$.  The Schwarzschild metric
is the final result of these manipulations.

\subsection{The event horizon}

Recall that our goal in this section is to understand how relevant the
Schwarzschild solution is as a description of physics, as opposed to
as a simple mathematical solution of the Einstein field equations.
Birkhoff's Theorem tells us that, at the very least, this solution
describes the exterior of at least some objects we might expect to
encounter in the universe.  However, what about for extremely compact
objects?  We raised concerns about oddness at the radius $r = 2M$.
Might there exist objects whose surfaces satisfy $r \le 2M$, and if
so, how do we understand the nature of the spacetime in that region?

Newtonian intuition {\cite{mitchell}} suggests that we should in fact
be concerned about the character of spacetime in this region.  The
escape speed from a spherical object of mass $M$ and radius $R$ is
found by requiring that the kinetic energy per unit mass equal the
work per unit mass it takes to move from $r = R$ to infinity:
\begin{equation}
\frac{1}{2}v_{\rm esc}^2 = \frac{GM}{R}\;.
\end{equation}
What is the radius at which the Newtonian escape speed is the speed of
light?  Plugging in $v_{\rm esc} = c$ and solving for $R$, we find
\begin{equation}
R = \frac{2GM}{c^2} = 2M\;.
\end{equation}
This is precisely the radius at which the Schwarzschild metric behaves
oddly!  Though this calculation abuses quite a few physical concepts,
it properly illustrates the fact that at this radius the impact of
gravity on radiation is likely to be extreme.

To understand what happens in GR, we need to examine ``stuff'' moving
around in this region\footnote{Note added Spring 2024: when this document was originally written, I apparently made an elementary error integrating the equations of motion to find these solutions presented in Eqs.\ (\ref{eq:geod_of_t}) and (\ref{eq:geod_of_tau}).  Reposting nearly 20 years later on the off-chance that someone might stumble on these formulas.}.  Consider dropping a particle from some
starting radius $r_0$ on a purely radial trajectory (no angular
momentum).  By integrating the geodesic equation, we find $r(t)$, the
position of the particle as a function of time measured by distant
observers, as well as $r(\tau)$, the position of the particle as a
function of time measured by the particle itself.  Setting $t = \tau =
0$ at the moment the particle is released we find
\begin{equation}
\frac{t}{2M} = \ln\left[\frac{\sqrt{\frac{r(r_0 - 2M)}{2M(r_0-r)} + 1}}{\sqrt{\frac{r(r_0 - 2M)}{2M(r_0-r)}} - 1}\right] + \sqrt{\frac{r(r_0 - r)}{(2M)^2}\left(\frac{r_0}{2M} - 1\right)} + \left(\frac{r_0}{2M} + 2\right)\sqrt{\frac{r_0}{2M} - 1}\left[\frac{\pi}{2} - \arctan\left(\sqrt{\frac{r}{r_0-r}}\right)\right]\;,
\label{eq:geod_of_t}
\end{equation}
\begin{equation}
\frac{\tau}{2M} = \left(\frac{r_0}{2M}\right)^{3/2}\arctan\left(\sqrt{\frac{r_0-r}{r}}\right) + \frac{\sqrt{rr_0(r_0 - r)}}{\sqrt{(2M)^3}}\;.
\label{eq:geod_of_tau}
\end{equation}
Expanding near $r = 2M + \delta r$, the behavior according to distant observers takes the asymptotic form 
\begin{equation}
\delta r = C(r_0)e^{-t/2M}\;.
\end{equation}
The quantity $C(r_0)$ we introduced here gather together several $r_0$ dependent factors
appearing in Eq.\ (\ref{eq:geod_of_t}) into the constant $C(r_0)$.  These two solutions for $r$ as a function of ``time'' are shown in
Fig.\ 1.
\begin{figure}
\includegraphics[width = 10cm]{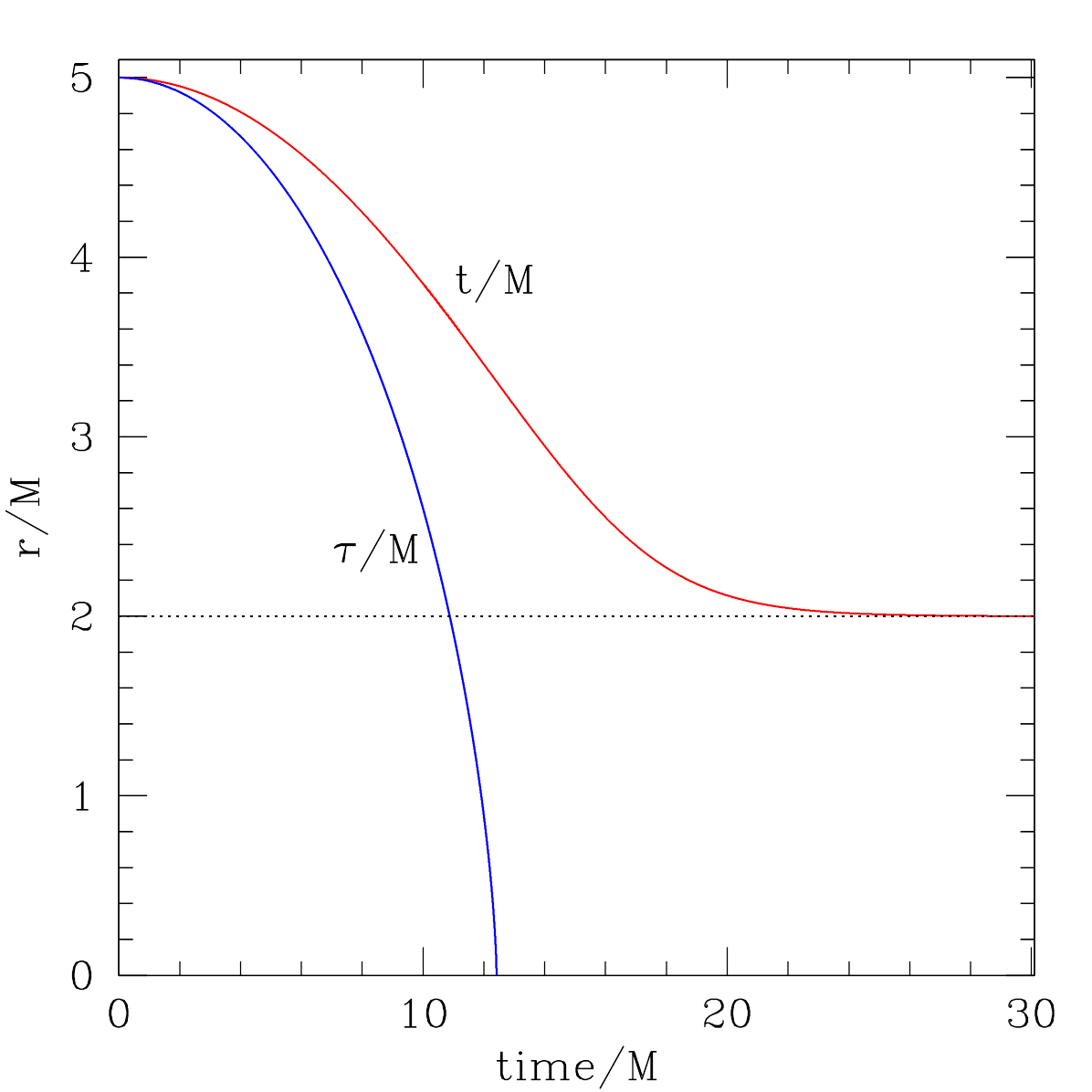}
\label{fig:geods}
\caption{The radial position of an infalling particle as a function of
two different measures of time: $t$, the time measured by distant
observers' clocks; and ``proper time'' $\tau$, time that is measured
by observers who ride with the particle.  The particle reaches $r = 0$
in finite proper time, but only asymptotically approaches $r = 2M$
according to our distant observers.}
\end{figure}

The two observers have {\it radically} different views.  The observer
who rides along with the infall finds that it takes finite time pass
through $r = 2M$.  By contrast, distant observers using $t$ as their
time label never actually see the particle fall through $r = 2M$ ---
the particle only asymptotically approaches this radius as $t \to
\infty$.  There appears to be a {\it severe} inconsistency here.

Before explaining the source of this apparent inconsistency, it's
worth examining what, if anything, the infalling body experiences at
$r = 2M$.  As already mentioned, the fundamental gravitational
``force'' in general relativity is subsumed by the notion of tides
that a body experiences.  These tides are given by tensors that
describe spacetime curvature; the most important is the Riemann
curvature tensor.  As measured by the freely falling particle, the
Riemann tensor's components are given by
\begin{equation}
R_{\alpha\beta\gamma\delta} = \frac{AM}{r^3}\;,
\label{eq:riemann}
\end{equation}
where $A = \pm 2, \pm 1, 0$ depending on specific index values.
Notice that there is nothing special about the curvature at $r = 2M$
--- the small body passes through the radius, feeling continually
increasing tidal forces, without any particular notification that $r =
2M$ has come and gone.  The curvature diverges as $r \to 0$ --- that
represents a true, infinite tidal force singularity (at least
classically).

The disagreement between the ``infalling view'' and the ``distant
observer view'' thus remains mysterious.  To clarify this mystery
somewhat, consider radial ``null geodesics'', the trajectories
followed by radiation:
\begin{eqnarray}
ds^2 = 0 \qquad \longrightarrow \qquad dt &=& \pm\frac{dr}{1 - 2M/r}
\\ &\equiv& \pm dr^*
\end{eqnarray}
where $r^* = r + 2M\ln(r / 2M - 1)$.  It takes an extremely long time,
according to distant observers, for light to leak out from near $2M$.
The delay becomes infinite if the light starts at $r = 2M$ ---
photons released at that radius {\it never} reach distant observers.
Consider next the energy associated with radiation moving on this
geodesic.  To do so, we use the rule that the measured energy of a
photon with 4-momentum ${\bf p}$ is given by
\begin{equation}
E = -{\bf p}\cdot{\bf u}\;,
\end{equation}
where ${\bf u}$ is the 4-velocity of the observer who makes the
measurement.  For a static observer at radius $r$,
\begin{equation}
u^a = \left[(1 - 2M/r)^{-1/2},0,0,0\right]\;.
\end{equation}
We imagine that a photon is emitted at radius $r$ and observed very
far away:
\begin{equation}
\frac{E_{\rm obs}}{E_{\rm emit}} = \sqrt{1 - \frac{2M}{r}}\;.
\end{equation}
Radiation emitted near $r = 2M$ is highly redshifted, with $r = 2M$
corresponding to a surface of infinite redshift\footnote{Recall that
redshift $z$ is defined via $E_{\rm obs}/E_{\rm emit} = 1/(1 + z)$.}.

Both the ``infinite time delay'' and the ``infinite redshift'' tell us
that our time coordinate is broken in the strong field of the black
hole.  The surface $r = 2M$ corresponds to a singularity in
Schwarzschild coordinates --- as seen by distant observers, clocks
slow to a halt as they approach this surface.  From far away, we never
actually see our test particle cross this surface; no paradox is
involved, however, since the photons we might use to observe this
behavior are infinitely redshifted away, and infinitely delayed in
reaching us.

\subsection{Other coordinate systems}

The punchline of the preceding section is that Schwarzschild
coordinates are not a good way to describe the kinematics of objects
that pass through the event horizon.  Fortunately, we can transform to
different coordinates which do a much better job of illustrating
aspects of motion in the strong field of the Schwarzschild spacetime.
For example ``Eddington-Finkelstein coordinates'' ({\cite{carroll}},
\S5.6; {\cite{mtw}}, Box 31.2) are very well adapted to describing
what would be observed by someone falling into a black hole; the
Painlev\'e-Gullstrand coordinate system ({\cite{poisson}}, \S5.1.4)
give a very nice description of the spacetime using coordinates
adapted to freely falling particles.  Here, we will focus almost
entirely on {\it Kruskal-Szekeres} (K-S) coordinates --- a (rather
nonintuitive) relabeling of time and radius that clarifies which
regions of spacetime are in causal contact with one another.

\subsubsection{Kruskal-Szekeres}

K-S coordinates $u$ and $v$ are related to Schwarzschild coordinates
$r$ and $t$ via
\begin{eqnarray}
u &=& \sqrt{1 - r/2M}e^{r/4M}\sinh(t/4M)
\\
v &=& \sqrt{1 - r/2M}e^{r/4M}\cosh(t/4M)\qquad r < 2M\;;
\end{eqnarray}
\begin{eqnarray}
u &=& \sqrt{r/2M - 1}e^{r/4M}\cosh(t/4M)
\\
v &=& \sqrt{r/2M - 1}e^{r/4M}\sinh(t/4M)\qquad r > 2M\;;
\end{eqnarray}
the opposite signs for $u$ and $v$ are also valid.  Notice that
\begin{equation}
u^2 - v^2 = (r/2M - 1)e^{r/2M}\;.
\end{equation}
This means that the surface $r = 2M$ is represented by the lines $u =
\pm v$, and the point $r = 0$ is represented by the hyperbolae $v =
\pm\sqrt{1 + u^2}$; these locations are plotted in Fig.\ 2.  The
right-hand quadrant of this figure represents $r > 2M$; the top
quadrant represents $r < 2M$.  For discussion of the meaning of the
left-hand and bottom quadrants, see {\cite{carroll}}, \S5.7;
{\cite{mtw}}, \S 31.5; or {\cite{poisson}}, \S5.1.  The seemingly
obtuse definitions of $u$ and $v$ in fact follow in a natural way from
Eddington-Finkelstein coordinates, as discussed in the
afore-referenced texts.

\begin{figure}
\includegraphics[width = 10cm]{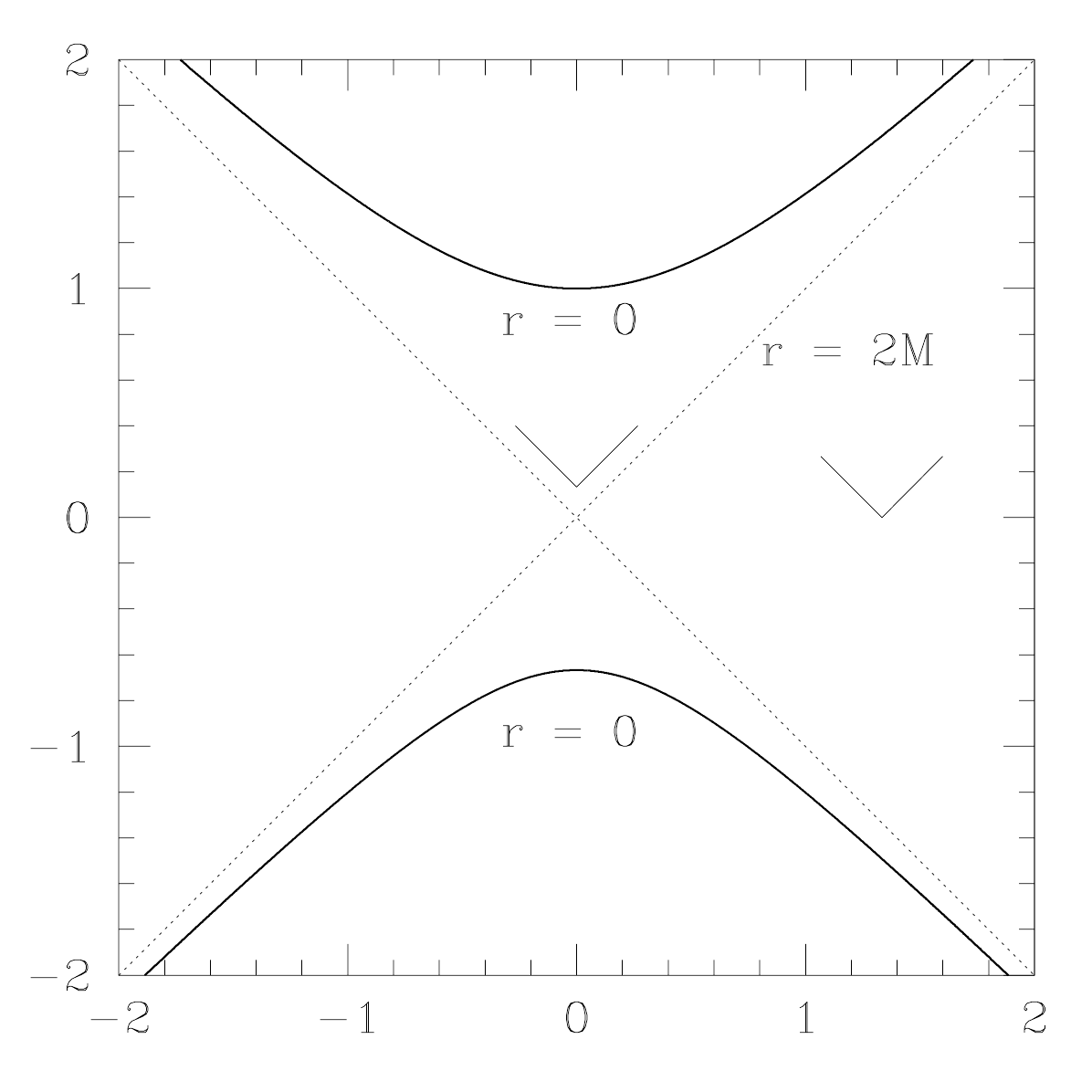}
\label{fig:kruskal}
\caption{The Schwarzschild spacetime in Kruskal-Szekeres coordinates,
and some example radial null geodesics.  The coordinate $u$ is
horizontal, $v$ is vertical.  The dotted lines represent the surface
$r = 2M$; the heavy black lines are the ``point'' $r = 0$.  The
smaller $45^\circ$ lines represent null geodesics, or radiation.
Notice that light directed radially outward never crosses $r = 2M$.
Notice also that any radiative trajectory that at any time crosses $r
= 2M$ {\it must} hit the $r = 0$ singularity at some point.}
\end{figure}

In terms of $u$ and $v$, the metric (\ref{eq:schw_metric}) becomes
\begin{equation}
ds^2 = \frac{32M^3}{r}e^{-r/2M}(-dv^2 + du^2) + r^2(d\theta^2 +
\sin^2\theta\,d\phi^2)\;.
\end{equation}
The virtue of these coordinates is made clear by considering radial
($d\theta = d\phi = 0$) null geodesics ($ds = 0$), describing light or
other radiation:
\begin{equation}
ds^2 = 0 \qquad \longrightarrow\qquad du = \pm dv\;.
\end{equation}
{\it Radiation travels on $45^\circ$ lines in K-S coordinates.}  A few
radiative trajectories are illustrated in Fig.\ 2; rays emitted from
the same point in opposite directions form a light cone with opening
angle $90^\circ$.

Consider next radial {\it timelike} trajectories, which are followed
by material objects and observers.  For these trajectories, we can put
$ds = d\tau$, the proper time measured along the trajectory itself.
We then find
\begin{equation}
-1 = \frac{32M^3}{r}e^{-r/2M}\left[-\left(\frac{dv}{d\tau}\right)^2
+\left(\frac{du}{d\tau}\right)^2\right]\;.
\end{equation}
Rearranging slightly, we have
\begin{equation}
\frac{r/32M^3}{(dv/d\tau)^2}e^{r/2M} = \left[1 -
\left(\frac{du}{dv}\right)^2\right]\;.
\end{equation}
Since the left-hand side of this equation is positive, we must have
\begin{equation}
\left|\frac{du}{dv}\right| < 1\;.
\end{equation}
In other words, material objects move such that their angle relative
to the $v$ axis is $< 45^\circ$.  A timelike trajectory from a
particular point must live inside that point's lightcone.

Bearing these features of motion in K-S coordinates in mind,
re-examine Fig.\ 2.  Notice that {\it nothing which is inside $r = 2M$
can ever come out!}  Consider a radial light ray emitted just inside
the $r = 2M$ line.  It either goes at a $45^\circ$ angle smack into
the $r = 0$ singularity; or, it skims slightly inside the $r = 2M$
line.  This second light ray can never cross $r = 2M$ --- that line
and the light ray are parallel.  Instead, after some finite time, it
will eventually smash into the $r = 0$ singularity --- even an
``outward'' directed light ray eventually hits the singularity if it
originates at $r < 2M$.  Since timelike trajectories must move at an
angle $< 45^\circ$ with respect to the $v$ axis, material objects and
observers inside $r = 2M$ are likewise guaranteed to smash into the
singularity.

By contrast, consider radial light rays emitted outside $r = 2M$.  The
ray directed toward the event horizon clearly hits the singularity;
however, the outward directed ray always remains outside $r = 2M$.  As
we'll show in the next section, it travels arbitrarily far away ---
``escaping'' the region of the black hole.

The punchline is that {\bf all} physical trajectories which begin
inside or cross $r = 2M$ must eventually hit the singularity at $r =
0$ --- even the trajectories of radiation.  Anything that goes inside
this region can {\it never} cross back into the exterior region.  For
this reason, the spherical surface $r = 2M$ is called the ``event
horizon'': Events which occur inside this horizon cannot ever impact
events on the outside.  Spacetime interior to the horizon is {\it
causally disconnected} from the rest of the universe.

\subsubsection{Penrose diagram}

Although it is somewhat beyond the original scope of these lectures,
it is hard to resist including the following brief discussion.  With
one further coordinate transformation, we can map infinitely distant
events in the spacetime to a finite coordinate label: put
\begin{eqnarray}
v + u = \tan\left[(\psi + \xi)/2\right]\;,
\\
v - u = \tan\left[(\psi - \xi)/2\right]\;.
\end{eqnarray}
The $u,v$ coordinate representation of Schwarzschild remaps to the
$\xi$ (horizontal axis) and $\psi$ (vertical axis) representation
shown in Fig.\ 3.  This way of showing Schwarzschild is called a
``Penrose diagram''.  The heavy black horizontal line is the $r = 0$
singularity; the $45^\circ$ dashed lines are the event horizon, $r =
2M$; the $45^\circ$ thin solid lines represent infinitely distant
parts of the spacetime.  Indeed, this diagram makes clear that there
are several different varieties of ``infinity'': Spacelike
trajectories in the infinite limit ($r \to \infty$, $t \to $ finite)
all accumulate at the single point $\psi = 0$, $\xi = \pi$ --- the
lower right-most point on the thin solid line.  Timelike trajectories
($t \to \infty$, $r \to $ finite) accumulate at $\psi = \pi/2$, $\xi =
\pi/2$ --- the upper left-most point on this line.  Null trajectories
($t \to \infty$, $r \to \infty$) reach various points on that line.
See {\cite{carroll,mtw,poisson}} for further discussion.

\begin{figure}
\includegraphics[width = 10cm]{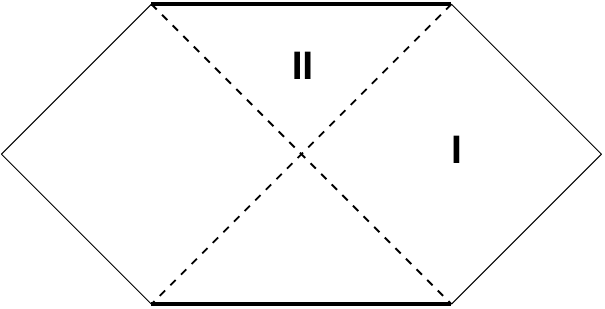}
\label{fig:penrose}
\caption{Penrose diagram of the Schwarzschild spacetime.  The heavy
black lines are the classical singularity at $r = 0$; the diagonal
dashed lines show the event horizon at $r = 2M$.  The region labeled
``I'' represents the exterior spacetime ($r > 2M$); that labeled ``I''
is the interior ($r < 2M$).}
\end{figure}

The line element $ds^2$ in the coordinates $\psi$, $\xi$ is given by
\begin{equation}
ds^2 = \frac{32M^3}{r}e^{-r/2M}\left(\frac{-d\psi^2 + d\xi^2}
{4\cos^2\left[(\psi+\xi)/2\right]\cos^2\left[(\psi-\xi)/2\right]}\right)
+ r^2(d\theta^2 + \sin^2\theta\,d\phi^2)\;.
\end{equation}
Just as in K-S coordinates, radiation travels on $45^\circ$ lines: the
condition for null trajectories, $ds = 0$, yields $d\xi = \pm d\psi$.
By examining light trajectories in the Penrose diagram, we see even
more clearly that any signal emitted at $r < 2M$ (region II of Fig.\
3) will forever be trapped inside.  Outward directed signals in region
I, by contrast, eventually reach ``infinity''.  Two observers in
region I can easily send signals to one another; they can also send
signals to a friend in region II.  However, that friend cannot send a
signal back out: region II cannot send any signal to region I.

\subsection{Summary of Schwarzschild spacetime}

We wrap up this section by summarizing what this brief examination of
the features of the Schwarzschild spacetime revealed:

\begin{itemize}

\item The Schwarzschild spacetime represents, in general relativity, a
monopolar ``gravitational field''.  One can regard it is the GR analog
of the Coulomb point charge electric field.

\item This solution describes the exterior spacetime of {\it any}
spherically symmetric body, even one that is time dependent (as long
as the time dependence preserves spherical symmetry).

\item The spacetime contains an {\it event horizon}: A spherical
surface at $r = 2M$ that causally disconnects all events at $r < 2M$
from those at $r > 2M$.  Things can go into the horizon (from $r > 2M$
to $r < 2M$), but they cannot get out; once inside, all causal
trajectories (timelike or null) take us inexorably into the classical
singularity at $r = 0$.

\item From the perspective of distant observers, dynamics near the
horizon appears weird --- such observers never actually see anything
cross it and fall in.  This is largely a consequence of extreme
redshifting --- clocks slow down so much, relative to distant
observers, that physical processes appear to come to a stop.  At any
rate, this weirdness is hidden from distant observers due to the
extreme redshifting and deflection of the photons that they would use
to observe these processes.

\end{itemize}

For all these reasons, the Schwarzschild solution is known as a {\it
black hole}: A region of spacetime that is completely dark and whose
interior is completely cut off from the rest of the universe.

\section{Dynamics: Creating a Schwarzschild black hole}
\label{sec:collapse}

The Schwarzschild solution that we have discussed so far is eternal
--- it has existed for all time.  Far more interesting are objects
that are created through astrophysical processes.  One may then ask
whether this seemingly special solution can actually be created from
``reasonable'' initial conditions.  In other words, can we make a
Schwarzschild black hole by starting out with something made from
``normal'' matter?

In an idealized but illustrative calculation, Oppenheimer and Snyder
{\cite{os39}} showed in 1939 that a black hole in fact does form in
the collapse of ordinary matter.  They considered a ``star''
constructed out of a pressureless ``dustball''.  By Birkhoff's Theorem,
the entire exterior of this dustball is given by the Schwarzschild
metric, Eq.\ (\ref{eq:schw_metric}).  Due to the self-gravity of this
``star'', it immediately begins to collapse.  Each mass element of the
pressureless star follows a geodesic trajectory toward the star's
center; as the collapse proceeds, the star's density increases and
more of the spacetime is described by the Schwarzschild metric.
Eventually, the surface passes through $r = 2M$.  At this point, the
Schwarzschild exterior includes an event horizon: A black hole has
formed.  Meanwhile, the matter which formerly constituted the star
continues collapsing to ever smaller radii.  In short order, {\it all}
of the original matter reaches $r = 0$ and is compressed
(classically!) into a singularity\footnote{Since all of the matter is
squashed into a point of zero size, this classical singularity must be
modified in a a complete, quantum description.  However, since all the
singular nastiness is hidden behind an event horizon where it is
causally disconnected from us, we need not worry about it (at least
for astrophysical black holes).}.  This evolution is illustrated in
Fig.\ 4.  Notice that the event horizon first appears at $t \simeq
6M$, and rapidly expands outwards.

\begin{figure}
\includegraphics[width = 10cm]{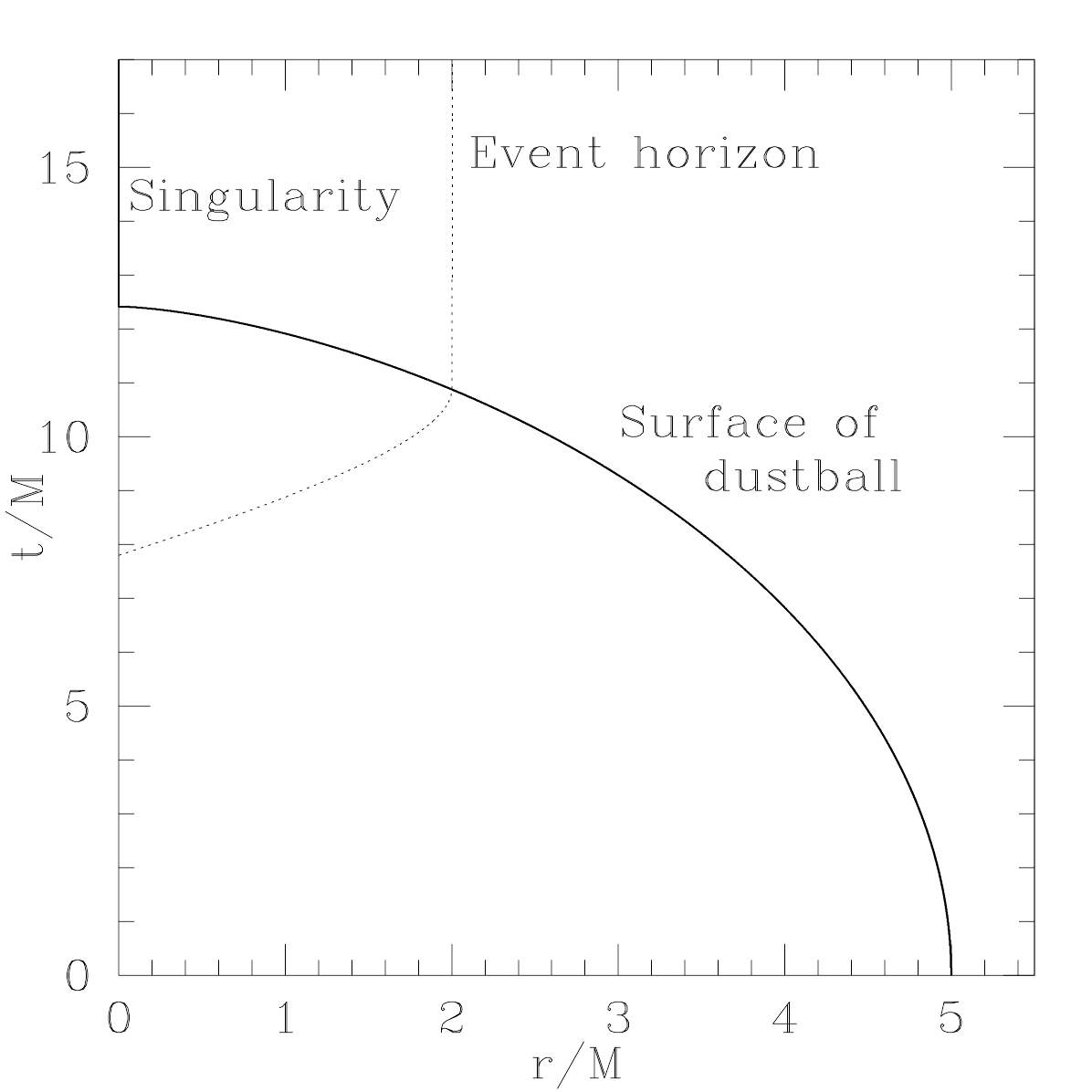}
\label{fig:os}
\caption{Oppenheimer-Snyder collapse of a ``star'' with an initial
radius $R_{\rm surf} = 5M$.  The surface of the star passes through $r
= 2M$ at a time of roughly $11M$; after that time, all of the star's
matter is behind the event horizon.  The matter continues collapsing,
eventually crushing into a zero-size singularity at roughly $t =
12.5M$.  The horizon actually first came into existence just before $t
= 6M$ and rapidly expanded outward.}
\end{figure}

The particularly attractive feature of the Oppenheimer-Snyder
construction is that it is completely analytic.  The metric which
describes the interior of a spherical dustball is
\begin{equation}
ds^2 = -d\tau^2 + a^2(\tau)\left[d\chi^2 + \sin^2\chi\left(d\theta^2 +
\sin^2\theta\,d\phi^2\right)\right]\;.
\end{equation}
This is the Friedmann-Robertson-Walker (FRW) metric; it is more
typically used to represent the large-scale spacetime of a
homogeneous, isotropic universe.  The radial coordinate inside the
star is $\chi$; it runs from the center ($\chi = 0$) to the star's
surface ($\chi = \chi_s$, to be determined later).  The function
$a(\tau)$ is the ``scale factor''; it provides an overall scale to
$\chi$.  The coordinate $\tau$ is proper time measured by a matter
element as it evolves with the dustball.

Enforcing Einstein's equations and applying appropriate initial
conditions (namely that each mass element is initially at rest, so
that the dustball collapses rather than expands) yields the following
parametric solutions for the proper time $\tau$ and the scale factor
$a$:
\begin{eqnarray}
a = \frac{a(0)}{2}\left(1 + \cos\eta\right)\;,
\label{eq:FRW_a}
\\
\tau = \frac{a(0)}{2}\left(\eta + \sin\eta\right)\;.
\label{eq:FRW_tau}
\end{eqnarray}
Collapse begins when the parameter $\eta = 0$; when $\eta = \pi$, all
of the matter has fallen into the singularity.

By Birkhoff's Theorem, the exterior spacetime is given by the
Schwarzschild metric.  The two descriptions must match smoothly at the
star's surface.  Matter on the star's surface can thus be described
either using the FRW coordinate description of Eqs.\ (\ref{eq:FRW_a})
and (\ref{eq:FRW_tau}), or by integrating the geodesic equation in
Schwarzschild coordinates.  Doing so the latter way yields the
following solutions for the radius of the star's surface, $R_{\rm
surf}$, and the elapsed time as measured by clocks at the surface,
$\tau_{\rm surf}$:
\begin{eqnarray}
R_{\rm surf} = \frac{R_{\rm surf}(0)}{2}\left(1 + \cos\eta\right)\;,
\label{eq:schw_r}\\
\tau_{\rm surf} = \sqrt{\frac{R_{\rm surf}(0)^3}{8M}}\left(\eta +
\sin\eta\right)\;.
\label{eq:schw_tau}
\end{eqnarray}
By computing the circumference of the star, ${\cal C} = \int
\sqrt{g_{\phi\phi}(\theta = \pi/2)}\,d\phi$, in the two descriptions,
we find
\begin{equation}
R_{\rm surf}(0) = a(0)\sin\chi_s\;.
\label{eq:circ_cond}
\end{equation}
By requiring that proper time at the surface be the same in the two
descriptions, we find
\begin{equation}
a(0) = \sqrt{\frac{R_{\rm surf}(0)^3}{2M}}\;.
\label{eq:tau_cond}
\end{equation}
Given an initial choice for the radius of the star, Eqs.\
(\ref{eq:circ_cond}) and (\ref{eq:tau_cond}) fix the values of $a(0)$
and $\chi_s$.  From there, the complete spacetime can be constructed.
This is how Fig.\ 4 was generated.

One might object to the relevance of this calculation --- after all,
pressureless matter isn't very interesting!  The point which this
calculation makes manifestly clear is that, at least in this idealized
case, one easily transitions from smooth, well-behaved initial data to
the rather odd black hole Schwarzschild solution.

Although, to our knowledge at least, there exist no similarly analytic
calculations of such collapse when for realistic matter (including
pressure, among other physically important characteristics), such a
calculation may be performed numerically; the first such calculation
was published in 1966 {\cite{mw66}}.  Although the quantitative
details are changed, the qualitative collapse picture is identical to
that given here: If the pressure is not sufficient to hold up the
infalling matter, the star's matter collapses until its surface passes
through $r = 2M$, leaving only an event horizon visible to external
observers.  Inside the star, the matter continues collapsing until it
is squashed into a zero-size singularity.

{\it Black holes are a generic end product of spherical gravitational
collapse.}

\section{The Kerr metric: Life is not spherical}
\label{sec:kerr}

In the limit of perfect spherical symmetry, the black hole picture is
now complete.  We have an exact solution that contains event horizons,
and we know that this solution can be produced from smooth,
well-behaved matter by gravitational collapse.  However, spherical
symmetry is a highly unrealistic limit --- we expect astrophysical
objects to have angular momentum (Schwarzschild black holes are
non-rotating), and to go through dynamical processes that are
extremely non-symmetric.  We cannot pretend that the perfectly
symmetric collapse process which produces Schwarzschild black holes
even approximates realistic astrophysical processes.

As we describe here and in the following section, these concerns are
addressed by what is called the {\it Kerr black hole}.  Discovered by
Roy Kerr in 1963 {\cite{kerr}}, this metric describes a rotating black
hole.  It is symmetric about the rotation axis, but not spherically
symmetric.  Furthermore, it can be shown that it is the {\it generic}
result of collapse to a black hole --- as we shall discuss in the next
section, once a black hole forms, it will shake and vibrate until its
metric settles down to the Kerr solution.

Written in ``Boyer-Lindquist'' coordinates {\cite{bl67}}, the Kerr
metric is
\begin{equation}
ds^2 = -\left(1 - \frac{2Mr}{\Sigma}\right)dt^2 - \frac{4 a M r
  \sin^2\theta}{\Sigma} dt\,d\phi + \frac{\Sigma}{\Delta}dr^2 +
  \Sigma\,d\theta^2 + \left(r^2 + a^2 +
  \frac{2Mra^2\sin^2\theta}{\Sigma}\right) \sin^2\theta\,d\phi^2\;,
\label{eq:kerr}
\end{equation}
where
\begin{eqnarray}
a &=& \frac{|\vec S|}{M} = \frac{G}{c}\frac{|\vec S|}{M}\;,
\\
\Delta &=& r^2 - 2Mr + a^2\;,
\\
\Sigma &=& r^2 + a^2\cos^2\theta\;.
\end{eqnarray}
In these coordinates, it is clear that the metric reduces to
Schwarzschild for $a = 0$.  It is also clear that this solution is
{\it not} spherical.  If it were spherical, we could find angular
coordinates $\theta',\phi'$ which are independent of $r$ and which
satisfy $g_{\phi'\phi'} = \sin^2\theta'\,g_{\theta'\theta'}$.  This
cannot be done for Kerr\footnote{Note that $g_{\phi\phi} =
\sin^2\theta\,g_{\theta\theta} + O(a^2)$; the metric is thus
approximately spherically symmetric for slow rotation.} except in the
Schwarzschild limit, $a = 0$.  The event horizon of the Kerr black
hole is located at $r_{\rm horiz} = M + \sqrt{M^2 - a^2}$, the larger
root of $\Delta = 0$.  (The smaller root corresponds to a horizon-like
surface inside the hole itself; this ``inner horizon'' plays no role
in astrophysical black hole physics, so we won't discuss it here.)

A particularly interesting feature of the Kerr spacetime is {\it frame
dragging}: The fact that $g_{\phi t} \propto a$ tells us that the
black hole spin ``connects'' time and space via the hole's rotation.
Geodesics in this spacetime show that objects are dragged by the
hole's rotation --- in essence, the gravitational field rotates,
dragging everything nearby to co-rotate with it.  Because of this,
Kerr is not a static spacetime, like Schwarzschild; the adjective used
to describe it is ``stationary''.  The spacetime does not vary with
time, but frame dragging makes it impossible for static observers to
exist near the black hole.  In the vicinity of a Schwarzschild black
hole, one can always arrange for an observer to sit in a static manner
relative to distant observers.  It may require an enormously powerful
rocket to prevent that observer from falling through the horizon; but
no issue of principle prevents one from using such a rocket.  By
contrast, close enough to the horizon of a Kerr black hole, {\it all}
observers are forced to co-rotate relative to very distant observers,
no matter how strongly they fight that tendency.

Unfortunately, though perhaps not surprisingly, there is no simple
collapse model akin to a rotating O-S collapse which demonstrates that
Kerr black holes are produced by collapse.  Fortunately, numerical
calculations due in fact show this; see for example Ref.\
{\cite{nakamura81}}.

The Kerr black hole is clearly far more relevant to astrophysical
discussions than the Schwarzschild solution --- it rotates, and it is
not spherically symmetric.  One might wonder: Are we done?  For
example, are there additional black hole solutions with no symmetry?
As we shall now discuss, the Kerr black hole is in fact the {\it
unique} black hole solution within general relativity.  Furthermore,
it is {\it stable}: If something momentarily disturbs the hole so that
the Kerr solution does not perfectly describe the spacetime, the hole
will vibrate momentarily and then settle back down to the Kerr
solution.  Taken together, these statements of uniqueness and
stability tell us that {\bf Kerr black holes are the ultimate outcome
of gravitational collapse in general relativity}.

\section{Uniqueness and stability of the black hole solution}
\label{sec:unique_stable}

The purpose of this section is to describe (without going into too
much technical detail) why the Kerr solution provides the {\it
complete} description of astrophysical black holes.  The first part of
this exercise is a description of the black hole uniqueness theorems.
They establish that {\it the Kerr black hole is the unique uncharged
black hole solution within general relativity}.  The uniqueness
theorems are known collectively as the ``no-hair'' theorem, meaning
that a black hole has no distinguishing features (no hair), except for
its mass and spin.  The next part is a discussion of stability and
Price's theorem, which establish the mechanism by which Nature
enforces the no-hair theorem.

Before going into this discussion, one might wonder: Should we
consider charged black holes?  For example, a non-spinning black hole
with mass $M$ and charge $Q$ is described by the Reissner-Nordstrom
metric {\cite{carroll,mtw,poisson}},
\begin{equation}
ds^2 = -\left(1 - \frac{2M}{r} + \frac{Q^2}{r^2}\right)dt^2 +
\frac{dr^2}{1 - 2M/r + Q^2/r^2} + r^2\left(d\theta^2 +
\sin^2\theta\,d\phi^2\right)\;;
\end{equation}
generalization to includes spin (the Kerr-Newman metric) is also
known.  Should we consider these black holes in our discussion?

At least on the rather long timescales relevant to measuring the
properties of black hole candidates, the answer to this question is
clearly {\it no}.  Typical astrophysical black holes are embedded in
environments that are rich in gas and plasma.  Any net charge will be
rapidly neutralized by the ambient plasma.  The time scale for this to
occur is very rapid, scaling roughly with the hole's mass: $T_{\rm
neut} \sim 5\,\mu{\rm sec}\,(M/M_\odot)$.  If charged black holes play
any role in astrophysics, it will only be on very short timescales ---
far shorter than is relevant for observations which probe the nature
of the black hole itself.

\subsection{``Black holes have no hair''}

The no-hair theorem tells us that (modulo the above discussion of
charge), {\it the only black hole solution is given by the Kerr
metric}.  A corollary of this is that black holes are {\it completely}
described by only two parameters, mass and spin.  {\bf No matter how
complicated the initial state, collapse to a black hole removes all
``hair'', leaving a simple final state.}

This is a remarkable statement.  Suppose that we begin with two
objects which are completely different --- a massive star near the end
of its life on one hand; a gigantic pile of elderly Winnebago campers
on the other.  We crush each object into a black hole such that each
has the same mass $M$ and spin $|\vec S|$.  The no-hair theorem tells
us that {\it there is no way to distinguish these objects}.  All
information about their highly different initial state is lost by the
time the event horizon forms.  This remarkable feature of classical
black holes is among their most bothersome; it is thus worth examining
this aspect with some care.

\subsubsection{Israel's theorem}

The first result which begins to establish the general no-hair
result was a theorem that was proved by Werner Israel in a 1967
publication {\cite{israel67}}.  Stated formally, Israel's theorem
proves that
\begin{quote}
Any spacetime which contains an event horizon and is static must be
spherically symmetric.  By Birkhoff's theorem it is therefore
described by the Schwarzschild metric.
\end{quote}
Hence, Schwarzschild black holes give the only {\it static} black hole
solutions.

The proof of this theorem is rather far beyond the scope of these
lectures; see {\cite{israel67}} for details.  A key element of the
proof is noting that the event horizon is a ``null surface'' --- it
can be thought of as ``generated'' by null (radiative) geodesics.
(This is demonstrated by Figs.\ 2 and 3, in which the event horizon
clearly corresponds to a piece of a light cone.)  The technical
details of the proof demonstrate that this null surface must be
spherically symmetric in static spacetimes if the spacetime is
asymptotically flat (more precisely, that spatial slices are
asymptotically Euclidean) and if the curvature is well-behaved over
the spacetime (i.e., it is smooth and bounded, with no singularities).

\subsubsection{Carter \& Robinson theorem}

Applying as it does only to static and spherically symmetric
spacetimes, Israel's theorem, though a landmark in understanding the
generic properties of black hole spacetimes, is of somewhat limited
applicability to astrophysics.  Fortunately, this theorem pointed the
way to its own generalization.  Work by Carter {\cite{carter71}} and
Robinson {\cite{robinson75}} showed that Israel's theorem could be
generalized from the static case as follows:
\begin{quote}
The only stationary spacetimes which contain event horizons are
described by the Kerr metric, with spin parameter $a \le M$.
\end{quote}
The detailed proof of this statement is, again, far beyond the scope
of these lectures; the interested reader is referred to Refs.\
{\cite{carter71,robinson75}} and references therein for the details.

\subsection{Enforcing ``no hair'': Price's theorem}

It may seem intuitively clear that the Israel-Carter-Robinson theorems
cannot possibly apply to, for example, newly born black holes that
form during supernova explosions; such situations are so highly
dynamical and asymmetric that there must be some ``structure''
(broadly speaking) that makes the spacetime more complicated than that
of the Kerr metric.  Fortunately, the no-hair theorem has a built in
loophole: It requires the spacetime to be {\it stationary}.  In other
words, the no-hair theorem requires that the spacetime itself be time
independent.  {\it Any} deviation from the pure black hole solution
means that this newly born black hole {\it cannot} be stationary!

This loophole leads us to what has come to be known as {\it Price's
theorem}: {\it Anything that can be radiated IS radiated.}  Deviations
from the ``perfect'' stationary black hole solution drive the emission
of gravitational radiation.  Backreaction from the that radiation
removes the deviation.  After a short time, the Kerr solution remains.
The statement ``black holes have no hair'' is more accurately given as
``black holes rapidly go bald''.

The mechanism by which the ``balding'' proceeds is most easily
explained for small deviations using black hole perturbation theory;
our discussion here follows that of Rezzolla {\cite{luciano}}.  We
begin by considering only perturbations to the Schwarzschild metric;
generalization to Kerr is conceptually straightforward, though (as we
outline briefly below) considerably more complicated calculationally.
Take the spacetime to be that of a Schwarzschild black hole plus a
small perturbation:
\begin{equation}
g_{ab} = g^{\rm BH}_{ab} + h_{ab}\;,\qquad ||h_{ab}||/||g^{\rm
BH}_{ab}|| \ll 1\;.
\end{equation}
The notation $||A_{ab}||$ means ``the typical magnitude of non-zero
components of the tensor $A_{ab}$''.  We now use this metric to
generate the Einstein tensor; since the black hole is a vacuum
solution, it satisfies $G_{ab} = 0$:
\begin{eqnarray}
G_{ab}\left[g^{\rm BH}_{ab} + h_{ab}\right] &\simeq&
G_{ab}\left[g^{\rm BH}_{ab}\right] + \delta G_{ab}\left[h_{ab}\right]
\\
&=& \delta G_{ab}\left[h_{ab}\right]\;.
\end{eqnarray}
The approximate equality on the first line follows from expanding the
Einstein equations to first order in the perturbation.  The equality
on the second line follows from the fact that the background black
hole metric is itself an exact vacuum solution.  This final result
tells us
\begin{equation}
\delta G_{ab}\left[h_{ab}\right] = 0\;.
\end{equation}
This gives us a wave-like operator acting on the perturbation.

Since the Schwarzschild background is static and spherically
symmetric, the metric perturbation can be expanded in Fourier modes
and spherical harmonics.  The ``time-time'' part of the perturbation
metric acts as a scalar function, and so it is expanded using ordinary
scalar spherical harmonics:
\begin{equation}
h_{00} = \sum_{lm} H^{lm}_0(t,r)Y_{lm}(\theta,\phi)\;.
\end{equation}
The ``time-space'' piece $h_{0i}$ can be regarded as a spatial vector,
and is thus expanded using vector harmonics:
\begin{equation}
h_{0i} = \sum_{lm}\left\{ H^{lm}_1(t,r) \left[Y^{\rm
E}_{lm}(\theta,\phi)\right]_i + H^{lm}_2(t,r) \left[Y^{\rm
B}_{lm}(\theta,\phi)\right]_i\right\}\;.
\end{equation}
The functions $[Y^{\rm E,B}_{lm}(\theta,\phi)]_i$ are electric- and
magnetic-type vector spherical harmonics.  These functions have
opposite parity: The electric harmonics are even parity, so $[Y^{\rm
E}_{lm}]_i \to (-1)^l [Y^{\rm E}_{lm}]_i$ for $\theta \to \pi -
\theta, \phi \to \phi + \pi$ (just like the usual scalar spherical
harmonic), while the odd parity magnetic harmonics obey $[Y^{\rm
B}_{lm}]_i \to (-1)^{l + 1}[Y^{\rm B}_{lm}]_i$.  The components of the
vector harmonics are given by ordinary spherical harmonics and their
derivatives; see {\cite{luciano}} for discussion tailored to this
application.

Finally, the ``space-space'' components of the metric perturbation
constitute a spatial tensor and are thus expanded in tensor harmonics:
\begin{equation}
h_{ij} = \sum_{lm}\left\{ H^{lm}_3(t,r)\left[Y^{\rm
pol}(\theta,\phi)\right]_{ij} + H_4(t,r) \left[Y^{\rm
ax}(\theta,\phi)\right]_{ij}\right\}\;.
\end{equation}
The ``polar'' modes are even parity; ``axial'' modes are odd.  The
angular functions again are given in terms of ordinary spherical
harmonics and their derivatives.

The metric perturbation is described in total by solving mode by mode,
for a particular choice of parity.  In each case, the gauge freedom of
general relativity allows us to set several of the functions $H_0 -
H_4$ to zero; running the remaining non-zero functions through the
wave-like equation $\delta G_{ab} = 0$ yields an equation of the form
\begin{equation}
\frac{\partial^2 Q}{\partial t^2} - \frac{\partial^2 Q}{\partial
r_*^2} + V(r) Q = 0\;.
\label{eq:schw_pert}
\end{equation}
The coordinate $r_* = r + 2M\ln(r/2M - 1)$; it asymptotes to the
normal Schwarzschild $r$ coordinate at large radius, but places the
event horizon in the limit $r_* \to -\infty$.  This facilitates
setting boundary conditions on the black hole.  The function $Q$ is
simply related to the non-zero $H$ functions; the potential $V$
depends on the parity of the mode.  At this point, we commonly insert
our Fourier decomposition so that the $H$ and $Q$ functions all
acquire a time dependence $\propto e^{-i\omega t}$.

The results for odd parity are particularly simple.  The coefficients
of the even parity harmonics are set to zero:
\begin{equation}
H_0 = H_1 = H_3 = 0\;.
\end{equation}
The function $Q$ relates to the remaining functions $H_2$ and $H_4$
via
\begin{eqnarray}
Q &=& \frac{H_4}{r}\left(1 - \frac{2M}{r}\right)\;,
\\
-i\omega H_2 &=& \frac{\partial}{\partial r_*}\left(r_* Q\right)\;.
\end{eqnarray}
The potential we find is
\begin{equation}
V(r) = \left(1 - \frac{2M}{r}\right)\left(\frac{l(l+1)}{r^2} -
\frac{6M}{r^3}\right)\;.
\label{eq:VRW}
\end{equation}
The equation governing odd-parity modes using the potential
(\ref{eq:VRW}) is known as the Regge-Wheeler equation {\cite{rw57}}.
A similar result, the Zerilli equation, governs even-parity modes
{\cite{z70}}.

To finally solve for these modes we must impose boundary conditions.
Since nothing can come out of the black hole, we require that the mode
be {\it purely ingoing} at the event horizon:
\begin{equation}
Q \propto e^{-i\omega(t + r_*)}\qquad r \to 2M\;.
\end{equation}
For similar reasons, we require that the mode be {\it purely outgoing}
at large radius:
\begin{equation}
Q \propto e^{-i\omega(t - r_*)}\qquad r \to \infty\;.
\end{equation}

Imposing these conditions, it is not too difficult to solve for $Q$
and thus construct the metric perturbation.  The result of this
exercise is that all modes {\it exponentially decay}, taking the form
of damped sinusoids.  The longest lived mode has frequency and decay
time
\begin{eqnarray}
\omega \simeq \frac{0.37}{M} \simeq 7.5 \times 10^4\,{\rm
sec}^{-1}(M_\odot/M)\;,
\\
\tau \simeq 1.72\,M \simeq 8.5\,\mu{\rm sec}\,(M/M_\odot)\;.
\end{eqnarray}
This tells us that {\it a distorted black hole settles down to the
``perfect'' black hole solution {\bf extremely} rapidly!}  This is the
mechanism by which the no-hair theorem is enforced: Any multipolar
distortion to the Schwarzschild metric is quickly shaken off.  On a
timescale of microseconds for stellar mass black holes, and seconds to
hours for the most massive black holes, the black hole relaxes to the
exact Schwarzschild solution.  It is to be emphasized that these are
{\it extremely} short astrophysical timescales --- any system that we
observe for a long period of time will certainly have relaxed into
this state.

Generalization of this argument to Kerr black holes is nontrivial, but
doable.  Largely due to the lack of spherical symmetry, expanding the
perturbed Einstein equations to find an equation governing the metric
perturbation does not yield separable equations.  It turns out,
though, that by developing a perturbation formalism based on expanding
curvature tensors rather than the metric one {\it can} develop
separable equations that are not too difficult to solve {\cite{t73}}.
Doing so, imposing similar boundary conditions, and solving the
resultant equations, one finds {\it exactly} the same qualitative
behavior.  The influence of spin causes some quantitative differences
(modes typically decay more slowly), but the general form of a
decaying sinusoid holds independent of spin.  {\it Objects with event
horizons rapidly settle down to the Kerr metric, radiating away all
distortions that push them from the ``perfect'' black hole solution.}

One might worry that, since these conclusions are based on
perturbative analyses, they are not relevant for ``large'' distortions
such as might be expected for newly born black holes in supernovae, or
the end result of black hole collisions.  Such worries are unfounded:
Detailed numerical analysis (by direct integration of the Einstein
field equations) confirms that this qualitative behavior holds even
for massive distortions from the quiescent black hole solution
{\cite{bhcollide}}.  One always finds that the distorted black hole
oscillates massive and violently, generating a large burst of
gravitational waves; but, those waves carry away the high-multipole
spacetime distortion.  A black hole described by the Kerr metric is
always what remains after the oscillations have damped out.

{\it Any black hole that forms in the universe should rapidly decay
into a form that is exactly described by the Kerr metric.}  Quiescent
black holes should be {\bf perfectly} described by an exact solution
of general relativity!

\section{Testing the black hole hypothesis}
\label{sec:test}

The discussion of the previous several sections establishes why, on
theoretical grounds, the black hole solutions of general relativity
are considered to be relevant to astrophysics.  Astrophysical
observations have now established many large, compact objects that are
obvious {\it candidate} black holes.  The big question becomes: Are
these black hole candidates in fact described by general relativity's
black hole solutions?  Several lines of investigation are working to
answer --- or at least constrain --- this question.

\subsection{Mass and compactness}

Many black hole candidates are observed as companions to normal stars
in binary systems.  If the companion is clearly compact and dark ---
observations establish that it is far smaller than a typical star
(several to several tens of kilometers), and that it is not luminous
--- then it is a good candidate to be a black hole.  However, it could
be a neutron star --- a fascinating and important object, but not the
subject of these lectures.

The first check to distinguish these two possibilities is to ``weigh''
the dark companion.  By studying orbits of binaries and applying
Kepler's laws, one can infer the mass of the dark companion (or at
least constrain it to some range).  If the object turns out to be
greater than about 3 solar masses, our current wisdom is that the
object is very likely to be a black hole --- to the best of our
knowledge, cold ``normal'' matter cannot hold itself up against
gravity at masses $\gtrsim 3\,M_\odot$ or so\footnote{The limit on the
maximum mass of a neutron star is not terribly well understood, as it
depends on the behavior of nuclear matter at extremely high density.
Our current best understanding of the nuclear equation of state at the
relevant densities suggests that the maximum allowed neutron star mass
is roughly $2 - 3\,M_\odot$ {\cite{kb96}}.}.  Observations establish
that the large dark object at the core of the Milky Way has a mass of
about $4\times 10^6\,M_\odot$ {\cite{sgrastar}}, and have established
that about 20 candidate black holes have masses around $5 -
20\,M_\odot$ {\cite{mr2004}}.

This argument relies on the fact that ``normal'' matter cannot hold
itself up against gravity at these high masses.  One can, however,
invoke ``non-normal'' matter --- condensations of scalar fields or
fermion balls, or ... This is in fact is what many of the posited
alternatives to the black hole hypothesis do.  To definitively check
the black hole hypothesis against these alternatives requires deeper
tests.

\subsection{Event horizons}

The defining property of black holes is their event horizon.  Rather
than a true surface, black holes have a ``one-way membrane'' through
which stuff can go in but cannot come out.  At least two tests seek to
test the black hole nature of black hole candidates by observationally
probing the existence of event horizons.

\subsubsection{X-ray bursts}

One test involves observations of x-ray binaries.  Such binaries
generically consist of a compact object in orbit with a ``normal''
star.  Gas flows from the star onto the compact object.  The gas is
heated during this accretion process to the point that it generates
x-rays.

In some of these sources, {\it x-ray bursts} are observed in addition
to the quiescent x-ray luminosity.  Bursts are now understood to arise
from thermonuclear detonation of accreted material
{\cite{b98,nh2003}}: Material accreted from the star builds up on the
surface of the compact object and is compressed by the object's
gravity.  After sufficient material accumulates, it undergoes unstable
thermonuclear burning, which we observe as an x-ray burst.

A key element of this model is that the compact object {\it must have
a surface}.  Material cannot accumulate on an event horizon, and so no
bursts can come from an x-ray binary whose compact object is a black
hole.

Observations show that {\it bursts are only seen if the compact
object's mass is sufficiently low that it is probably a neutron star}
{\cite{narayan}}.  Bursts are {\it never} seen from sources whose mass
are high enough that they are black hole candidates.  This is totally
consistent with the idea that black holes lack hard surfaces on which
material can accumulate.  See {\cite{narayan}} for additional
discussion of this idea and further details.

\subsubsection{Imaging a black hole's shadow}

The black hole candidate which covers the largest fraction of our sky
is the one which resides in the center of our galaxy.  This object has
a mass of about $4\times 10^6\,M_\odot$ {\cite{sgrastar}}; if it is
indeed a black hole, this corresponds to a horizon radius of about $6
- 12$ million km.  The galactic center is about 7000 parsecs away, so
the horizon subtends an angle of about 0.01 milliarcseconds on the
sky.  This is not too far from what might be achievable in the near
future using very long baseline radio interferometry.  Indeed, recent
observations have shown that this source must have a size smaller than
about 1 astronomical unit {\cite{shenetal}}; the precision required
for this measurement is within an order of magnitude or so of what is
needed to probe the region near the event horizon.

What we really want to measure is the horizon's {\it shadow}: the
``hole'' cast in background radiation due to the presence of the event
horizon.  Because of the bending and focusing of radiation in the
strong gravitational field of a black hole, the shadow will be
somewhat larger than the event horizon.  For a Schwarzschild black
hole, the shadow would be a perfect circle $3\sqrt{3}/2 \simeq 2.6$
times larger than the event horizon.  For rotating black holes (the
generic case which we expect), the shadow is expected to be somewhat
asymmetric due to frame dragging --- the bending of light is not
symmetric.  Getting the resolution required to do such imaging will
require interferometry on the scale of the Earth's size, observing at
sub-mm wavelengths; such a measurement is plausible in the near
future.  This idea is discussed in detail in Ref.\ {\cite{fma2000}}.

\subsection{Orbits}

Finally, there is great potential in probing the spacetimes of black
hole candidates by observing orbits about them (much as we examined
``thought orbits'' to probe the nature of the event horizon in Sec.\
{\ref{sec:schw}}).  If orbits can be measured precisely, then we can
measure informative characteristics like the timescales associated
with different motions.  These can be be connected to a model of the
spacetime, from which the spacetime's character can be inferred.  Such
measurements have already begun by studying orbits around the black
hole at the core of the Milky Way {\cite{sgrastar}}.  Those stars are
in orbits which, strictly speaking, a relativist would consider weak
field\footnote{It should be noted that one star moves at about 5000
km/sec through the periapsis of its orbit.  Though formally weak
field, these orbits are still rather extreme!}, but they are
constrained well enough that they make it possible to determine the
putative black hole's mass with good accuracy.

In general, motion around a black hole is described by integrating the
geodesic equation for the black hole's spacetime.  The orbits admit a
notion of conserved energy $E$, of conserved ``axial'' angular
momentum $L_z$, as well as a third integral $Q$ which is essentially
the ``rest'' of the orbit's angular momentum: $Q = |\vec L|^2 - L_z^2$
for non-spinning black holes; for spinning black holes, the
interpretation is more complicated, though this intuitive picture of
$Q$ is useful.  Using these conserved constants, the geodesic
equations can be rewritten in terms of oscillations in the $r$ and
$\theta$ coordinates, while the orbits ``whirls'' with respect to
$\phi$:
\begin{eqnarray}
\Sigma^2\left(\frac{dr}{d\tau}\right)^2 &=& \left[E(r^2+a^2)
- a L_z\right]^2- \Delta\left[r^2 + (L_z - a E)^2 +
Q\right] \equiv R(r)\;,
\\
\Sigma^2\left(\frac{d\theta}{d\tau}\right)^2 &=& Q - \cot^2\theta L_z^2
-a^2\cos^2\theta(1 - E^2) \equiv \Theta(\theta)\;,
\\
\Sigma\left(\frac{d\phi}{d\tau}\right) &=&
\csc^2\theta L_z + aE\left(\frac{r^2+a^2}{\Delta} - 1\right) -
\frac{a^2L_z}{\Delta} \equiv \Phi(r,\theta)\;,
\\
\Sigma\left(\frac{dt}{d\tau}\right) &=&
E\left[\frac{(r^2+a^2)^2}{\Delta} - a^2\sin^2\theta\right] +
aL_z\left(1 - \frac{r^2+a^2}{\Delta}\right) \equiv T(r,\theta)\;;
\\
\Sigma &=& r^2 + a^2\cos^2\theta\qquad \Delta = r^2 - 2Mr + a^2\;.
\end{eqnarray}
One feature of these orbits is the existence of ``innermost stable
orbits''.  For example, careful analysis of the radial potential
$R(r)$ for $a = 0$ shows that no stable circular orbits can exist at
$r < 6M$ --- small perturbations drive such orbits to plunge almost
immediately into the hole's event horizon.  Another feature is that
the familiar Keplerian orbital period, $T = 2\pi\sqrt{a^3/M}$ (where
$a$ is orbital semi-major axis) is split into 3 distinct periods,
corresponding to motions in $r$, $\theta$, and $\phi$.  These periods
are shown in Fig.\ 5; notice that they asymptote to the Kepler result
far from the black hole, but are strongly split in the strong field.
The distinct character of these orbital timescales is a fingerprint of
the black hole strong field.

\begin{figure}
\includegraphics[width = 10cm]{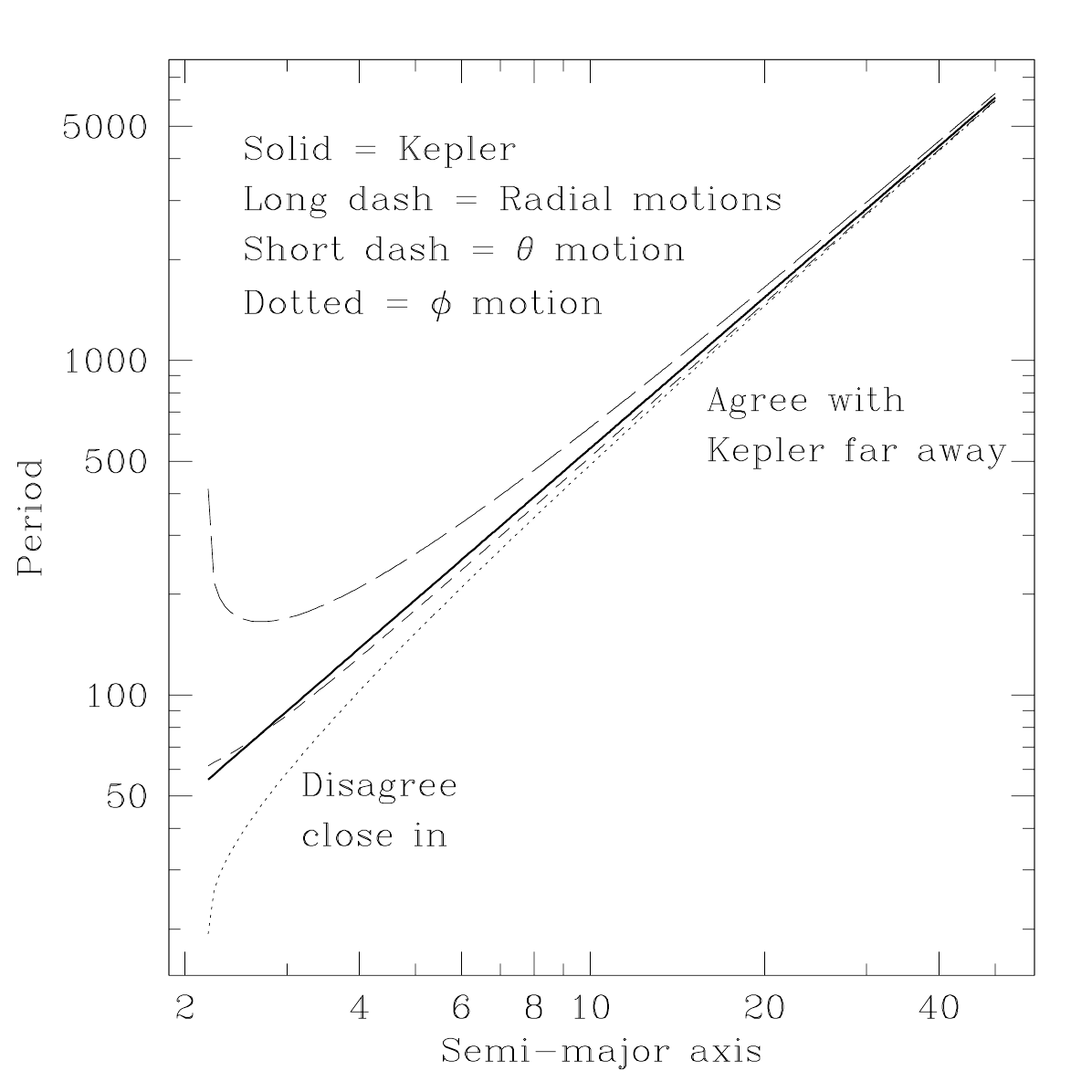}
\label{fig:periods}
\caption{Orbital periods around a rotating black hole.  Notice that
they asymptotically approach the Kepler result for large semi-major
axis, but are distinctly split in the strong field.}
\end{figure}

At least two types of measurements have the promise to probe orbits in
the strong field and to observe these distinct behaviors: x-ray
measurements of accretion (the flow of gas onto the black hole in
these potentials), and gravitational waves (arising from the motion of
bodies orbiting in these potentials).  A particularly interesting
strong-field hallmark visible in some accreting systems is a
fluorescence line associated with iron, the Fe K$\alpha$ line.  In the
rest frame, this line is expected to be a sharp feature at an energy
of $6.4$ keV.  The observed lines are highly broadened from $4$ to $7$
keV --- both blueshifted and redshifted due to a combination of the
gravitational redshift and the rapid whirling motion of gas near the
black hole.  With some assumptions about the nature and structure of
the gaseous disk from the radiation originates, measurement of such
lines makes it possible to estimate the spin of massive black holes
{\cite{rn2003}}; clean, high signal-to-noise data may make it possible
to constrain the spacetime even more sharply, perhaps enabling a
careful test of the black hole hypothesis.

Future gravitational wave observations likewise promise to enable
careful studies of black hole spacetimes.  Gravitational waves are
generated by a source with a varying mass quadrupole moment; binary
systems are a perfect example of a strong gravitational radiator.  It
is expected that galactic black holes will occasionally capture a
smaller compact object ($\sim 10-100\,M_\odot$ black hole or $\sim
1\,M_\odot$ neutron star) {\cite{gair}}.  That compact object will
generate gravitational waves with spectral content at all of the
orbital frequencies, $\Omega_r$, $\Omega_\theta$, and $\Omega_\phi$,
as well as harmonics of those frequencies.  Indeed, those frequencies
will slowly evolve as the small body loses orbital energy and angular
momentum due to the waves' backreaction; by tracking the evolution, a
great deal of information can in principle be extracted.  Future
space-based gravitational-wave missions (see Larson's contribution to
this summer school) are expected to measure these waves well enough to
precisely determine the masses and spins of black holes
{\cite{bc2004}}, and should be able to check whether the spacetime
satisfies the stringent requirements of the black hole hypothesis
{\cite{multipoles}}.

\section{Conclusions}

Unambiguous observational evidence for the existence of astrophysical
black holes has not yet been established.  The moral of classical
general relativity is that black holes must be created in the right
circumstances; and, they must be driven to a particularly simple,
unique form.  Chandraskehar {\cite{chandra}} provides a typically
clear summary:
\begin{quote}
It is well known that the Kerr solution with two parameters provides
the unique solution for stationary black holes that can occur in the
astronomical universe.  {\it But a confirmation of the metric of the
Kerr spacetime (or some aspect of it) cannot even be contemplated in
the foreseeable future.}
\end{quote}
(Emphasis added.)  Chandra's final sentence makes sense in the context
of 1986, when this statement was made.  Technology has advanced
considerably since then --- as observations continue to improve, the
amount of wiggle room in theoretical phase space for alternatives to
the black hole hypothesis grows progressively smaller and smaller.  We
already are gathering evidence allowing confirmation of aspects of
black hole spacetimes; it is likely that definitive, unambiguous
evidence showing --- or disproving! --- that the many black hole
candidates we see today are general relativity's black holes is now
not too far in the future.

\begin{acknowledgments}
I wish to thank the organizations of the 2005 SLAC Summer Institute
for giving me the opportunity to assemble these lectures, and for
their patience as I wore out deadlines completing this final writeup.
My work on black hole physics is supported by NSF Grants PHY-0244424
and CAREER Grant PHY-0449884, and by NASA Grants NAG5-12906 and
NNG05G105G.  I also very gratefully acknowledge the support of MIT's
Class of 1956 Career Development Fund.
\end{acknowledgments}

\end{document}